\documentclass[aps,pra,twocolumn,superscriptaddress]{revtex4-1}
\usepackage{graphicx}
\usepackage{ulem}
\usepackage{dcolumn}
\usepackage{bm}
\usepackage{amsmath}
\usepackage{amssymb}
\usepackage{amsfonts}%
\usepackage{color}
\usepackage{hyperref}
\begin{document}

\title{Quantum tunneling isotope exchange reaction H\texorpdfstring{$_2$}{} + D\texorpdfstring{$^- \rightarrow$}{}  HD + H\texorpdfstring{$^-$}{}}
\author{Chi Hong Yuen}
\affiliation{Department of Physics, University of Central Florida, Orlando, Florida 32816, USA}
\author{Mehdi Ayouz}
\affiliation{Laboratoire de Genie des Procedes et des Materiaux, Ecole Centrale Paris, CentraleSupelec,campus de Chatenay-Malabry Grande Voie des Vignes F-92 295 Chatenay-Malabry Cedex, France}
\author{Eric S. Endres}
\author{Olga Lakhmanskaya}
\author{Roland Wester}
\affiliation{Institut f\"{u}r Ionenphysik und Angewandte Physik, Universit\"{a}t Innsbruck, Technikerstra\ss{}e 25, 6020 Innsbruck, Austria}
\author{Viatcheslav Kokoouline}
\affiliation{Department of Physics, University of Central Florida, Orlando, Florida 32816, USA}

\begin{abstract} 
The tunneling reaction H$_2$ + D$^-$ $\rightarrow$ HD + H$^-$ was studied in a recent experimental work at low temperatures (10, 19, and 23~K) by Endres {\it et al.}, Phys. Rev. A {\bf 95}, 022706 (2017). An upper limit of the rate coefficient was found to be about 10$^{-18}$ cm$^3$/s. In the present study, reaction probabilities are determined using the ABC program developed by Skouteris {\it et al.}, Comput. Phys. Commun. {\bf 133}, 128 (2000). The probabilities for ortho-H$_2$ and para-H$_2$ in their ground rovibrational states are obtained numerically at collision energies above 50~meV with the total angular momentum $J$ = 0 - 15 and extrapolated below 50~meV using a WKB approach. Thermally averaged rate coefficients for ortho- and para-H$_2$ are obtained; the largest one, for ortho-H$_2$ is about $3.1\times10^{-20}$ cm$^3$/s, which agrees with the experimental results.
\end{abstract}

\pacs{} 

\maketitle

\section{Introduction}
In the interstellar medium (ISM), the H$^-$ ion could be responsible for the formation of molecular anions, several of which have recently been observed  \cite{mccarthy06,gupta07,kentarou07,cernicharo07,cernicharo08,thaddeus08,herbst08,harada08,agundez10}. For example, the CN$^-$ ion \cite{agundez10} could be formed in the reaction H$^-$ + HCN $\to$ H$_2$+CN$^-$ \cite{douguet13,satta2015quantum}. However, H$^-$ has never been observed in the ISM by photoabsorption spectroscopy. Its detection is difficult because it has only one bound electronic state. A few unsuccessful efforts were made to search for it in the ISM, such as using far-ultraviolet autodetachment transitions \cite{ross08}. The existence of H$^-$ can also be inferred indirectly via the spectroscopy of the H$_3^-$ ion, which could be formed by radiative association of H$_2$ and H$^-$ in diffuse molecular clouds \cite{ayouz11}. 

For an eventual indirect detection of H$^-$ in the ISM via detection of H$_3^-$, studying the structure and formation of the H$_3^-$ system with its isotopologues is important \cite{ayouz10,ayouz11}. Recently, reactive scattering of the D$^-$ ion in collisions with H$_2$,
\begin{align}
\mathrm{H_2} + \mathrm{D^-} \rightarrow \mathrm{HD} + \mathrm{H^-}\,,
\label{eq:reaction}
\end{align}
were studied in an experiment by Endres {\it et al.} \cite{Endres2017}.  This exothermic reaction has a reaction barrier of 330 $\pm$ 60 meV \cite{haufler97}, which cannot be overcome by thermal activation at low temperatures. Therefore, at low temperatures in cold molecular clouds in the ISM, the reaction can proceed only by quantum tunneling and cannot be described with the classical Langevin theory. As H$_2$D$^-$ is one of the simplest anionic triatomic molecules, the reaction could serve as a benchmark process for quantum tunneling at low temperatures. The experiment of Ref.~\cite{Endres2017} was performed in a cryogenic 22-pole ion trap. From D$^-$ loss rate measurements, it was concluded that the upper limit of the rate coefficient at 10 K is 2.6 $\times$ 10$^{-18}$ cm$^3$/s. The absence of the H$^-$ signal gave a smaller upper limit of 9 $\times$ 10$^{-19}$ cm$^3$/s. 

\begin{figure}[h]
\centering
\includegraphics[width=7cm]{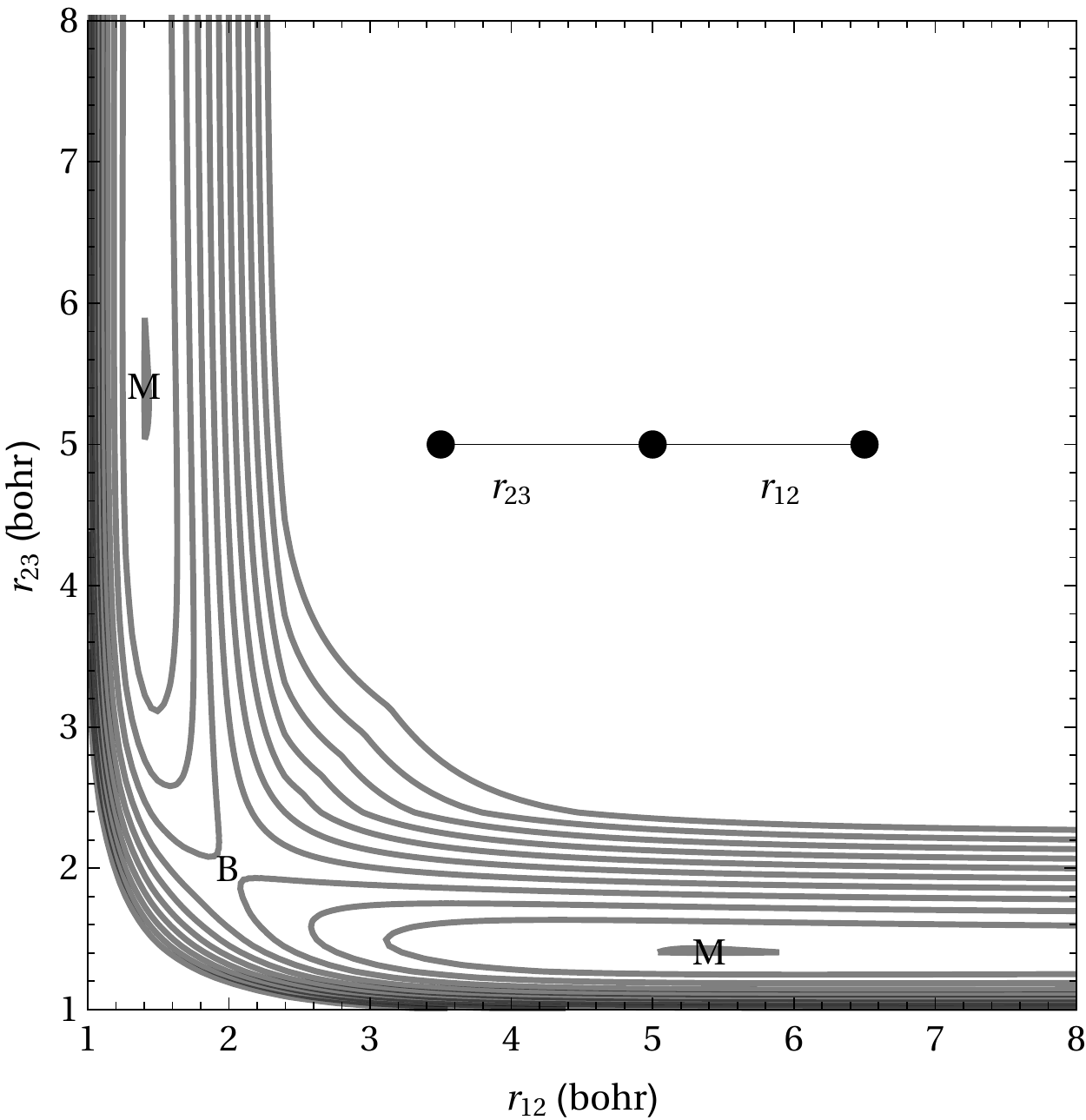}
\caption{\label{fig:potbarrier}H$_3^-$/H$_2$D$^-$ PES as a function of two internuclear distances $r_{12}$ and $r_{23}$ for linear geometries. The absolute energy minimum of the PES is indicated with M, and the peak of the potential barrier is labeled with B. Successive energy contours differ by 0.16 eV.}

\end{figure}

There have been several theoretical investigations of the reaction (\ref{eq:reaction}). However, the majority of the studies considered collision energies from about 0.3~eV to 2~eV. For example, a time-dependent wave packet method with and without Coriolis coupling was used in Refs.~\cite{morari2005,yao2006} and a quasi-classical trajectory method in Ref.~\cite{zhang2010}. The cross section for the process using both time-dependent and time-independent method was reported by Giri and Sathyamurthy \cite{giri2006}.  A comparison of reaction cross sections obtained using different available potential energy surfaces was made at collision energies above 0.3 eV in Ref.~\cite{wang2013}. There is also a study using the variational transition-state method, which gave the rate coefficient of the order of 10$^{-23}$ cm$^3$/s at 30 K \cite{luo2011}. At ultracold and cold temperatures, a similar reaction $\mathrm{H_2} (v = 0 - 5, j = 0, 1)+ \mathrm{D} \rightarrow \mathrm{HD} + \mathrm{H}$ was reported by Simbotin and C{\^o}t{\'e} \cite{simbotin2015}. Giri and Sathyamurthy  have also reported  \cite{giri07} theoretical results for the H$^-+$HD collisions with the HD molecule being in the first excited vibrational level. 
The present study is focused on a quantum-mechanical approach to obtain the cross section of reaction \eqref{eq:reaction} at low energies, below the potential barrier.

\section{Theoretical approach}

\begin{figure}[ht]
\centering
\includegraphics[width=7cm]{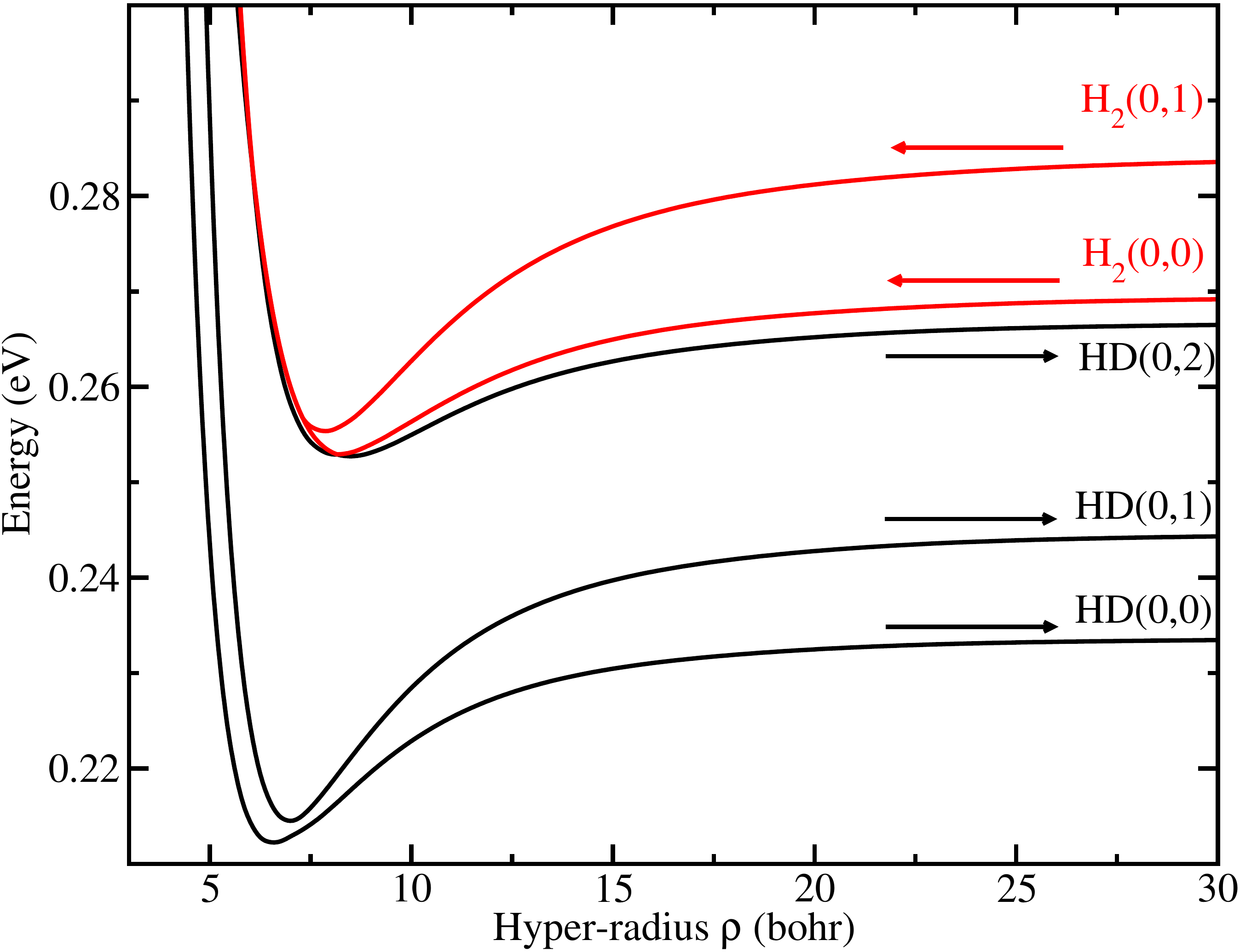}
\caption{Hyperspherical adiabatic potential energy curves for H$_2$D$^-$. At low collision energies between H$_2$($v$=0, $j$=0,1) and D$^-$, there are three open exit channels for HD($v,j$) with $v = 0$ and $j$ = 0, 1, 2 \cite{ayouz2013resonant}.}
\label{fig:adapot}
\end{figure}

In this study, we employ the H$_3^-$ potential energy surface (PES) calculated by Ayouz {\it et al.} \cite{ayouz10}. The PES has a barrier of about 450~meV above the dissociation limit of H$^-$ + H$_2$, when H$_2$ is at the the equilibrium geometry, along the minimum energy pathway of the reaction \eqref{eq:reaction} (see Fig.~\ref{fig:potbarrier}). The barrier is about two orders of magnitude larger than a typical collision energy at 10~K in the experiment by Endres {\it et al.} \cite{Endres2017}. Therefore, the reactive scattering is highly suppressed at 10~K. Figure~\ref{fig:adapot} shows the hyperspherical adiabatic (HSA) energy curves of H$_2$D$^-$ as a function of the hyper-radius (for details, see Refs.~\cite{ayouz2013resonant}). As Fig.~\ref{fig:adapot} demonstrates, at low energies of collisions between D$^-$ and the H$_2$ molecule in the rotational level $j=$ 0 or 1, the possible reaction channels are H$^-$+HD with HD being in the rotational state $j$=0, 1, or 2.

At low collision energies the reaction probability is extremely small, which requires a theoretical approach able to provide the required accuracy for the probabilities. To assess  and compare the accuracy of different numerical methods, we have performed a restricted calculation of the reaction probabilities for para-hydrogen in the ground rovibrational state with the total angular momentum $J=0$ using three different approaches: the time-independent hyperspherical adiabatic approach \cite{Yuen2017}, the multi-configuration time-dependent Hartree (MCTDH) method \cite{beck2000}, and the ABC program \cite{Skouteris2000}. Only a brief overview of the three methods is provided below, details can be found in \cite{Yuen2017,beck2000,Skouteris2000} and the references therein. 

The HSA approach represents different arrangements of the system by using a single set of hyperspherical coordinates. As a first step in this method one calculates HSA energies and wave functions at many fixed hyper-radii. At large values of the hyper-radius the energies and wave functions represent the rovibrational states of the dimer, as shown in Fig.~\ref{fig:adapot}. The non-adiabatic couplings between the HSA channels are represented using the modified slow variable discretization \cite{tolstikhin96}. Using the eigenchannel R-matrix approach, the scattering matrix is extracted at the last grid point of the hyper-radius without performing a coordinate transformation as is performed in Refs.~\cite{Pack1987,launay1989}. 
The MCTDH method is a general algorithm for solving the time-dependent  Schr\"{o}dinger equation for multi-dimensional systems. It has been successful in treating reactive scattering for different systems \cite{sukiasyan2001,wu2004}. This method propagates the incoming wave packet along the PES. The outgoing flux is absorbed by placing a complex absorbing potential at the reaction coordinate, and the reaction probability is extracted from the outgoing flux. 
The ABC program is also able to represent the reactive scattering of a dimer and an atom. It has been successfully applied to many three-body systems \cite{balakrishnan2001,bodo2002,simbotin2015}. It is based on solving the Schr\"{o}dinger equation using the coupled-channel approach in Delves hyperspherical coordinates. The coupled-channel basis functions are constructed on the surface of the hypersphere at each hyper-radius. Finally, the parity-adapted scattering matrix are obtained by applying boundary conditions.

\begin{figure}[ht]
\centering
\includegraphics[width=8cm]{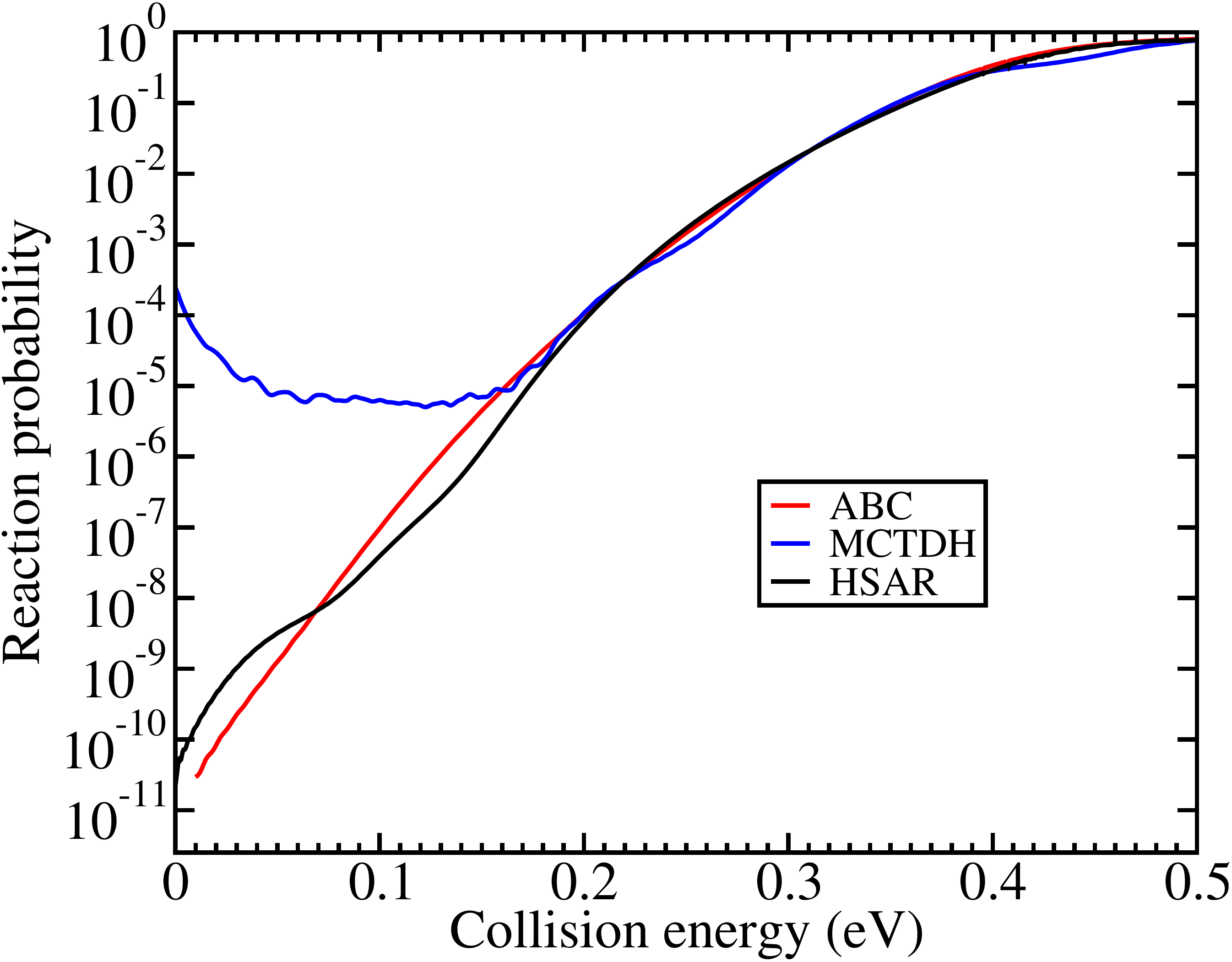}
\caption{(Color online) Reaction probabilities  of the reaction H$_2(0,0)$+D$^- \to$ HD+H$^-$ with the total angular momentum $J=0$, obtained from the HSA approach (black), the MCTDH method (blue) and the ABC program (red). }
\label{fig:compare}
\end{figure}

Figure~\ref{fig:compare} compares the results obtained using the three methods. At collision energies above 0.2 eV, the reaction probabilities from all three methods are in good agreement. Below that energy, in the tunneling regime, the results diverge. In the MCTDH calculation below 0.1 eV, the outgoing flux is extremely small, which sets a limit on accuracy of the calculation. To achieve a better accuracy in the MCTDH approach, one needs to increase significantly the propagation time and the length of the grid, which makes the calculation much more expensive compared to the ABC method. 
In the spectrum of the H$_2$+D$^-$ collisions, one expects to observe rovibrational resonances similar to the ones computed in Ref.~\cite{ayouz2013resonant}. Such resonances are situated at collision energies above 25 meV. Indeed, in the HSA spectrum shown in Fig.~\ref{fig:compare}, the peaks at energies about 0.1 eV and 0.15 eV correspond to the two series of resonances predicted in \cite{ayouz2013resonant}. At energies below 0.1 eV, the HSA results show the oscillations that do not correspond to any resonance. These are artifacts of the present HSA method, which is currently unable to represent very small tunneling probabilities below 0.1~eV. 
A disadvantage of the ABC program is that it does not represent rovibrational Feshbach resonances due to the restricted basis of coupled-channel functions employed in the approach. Because we are interested only in the thermally averaged rate coefficient at temperature about 10 K, the resonances structure at higher collision energies are irrelevant. Therefore, numerical stability of the method at low energies is the most important factor. At energies below 0.1 eV, the ABC program produced smooth reaction probability as expected since there is no resonance present in this region. Therefore, the ABC program was chosen to perform calculations for all values of $J$ needed to be included in the study.

In the ABC program, the total reaction cross section is obtained as
\begin{align}
  \sigma^P_{\lambda' \leftarrow \lambda v j}(E_i) = \frac{\pi}{k_i^2} \sum_{J} (2J+1) P_J(E_i), \nonumber\, \\
  P_J(E_i) =  \frac{1}{2j+1} \sum_{v'j'} \sum_{k'} \sum_{k}  |S^{J,P}_{\lambda'v'j'k' \leftarrow \lambda v j k}|^2,
  \label{eq:rate}
\end{align}
where $E_i$ is the collision energy; $k_i=\sqrt{2 \mu E_i}/ \hbar$ with $\mu$ being the reduced mass of the D$^-$ ion and the H$_2$ molecule; $\lambda$ and $\lambda'$ are the initial, H$_2$+D$^-$, and the final, HD+H$^-$, rearrangements correspondingly;  $J$ and $k$ are the total angular momentum and its projection on  a fixed axis in the molecular reference frame; $v,j$ and $v', j'$ are initial and final rovibrational levels of the H$_2$ and HD molecules, respectively; $P$ is the parity quantum number. In the above equation, $P_J$ has a meaning of the reaction probability for given values of $J$, $\lambda$, $\lambda'$, $v$, and $j$.

At collision energies above 50 meV, the cross section converges when channels with energies up to 2.3 eV above the dissociation limit and 16 or 17 rotational states for para- or ortho-H$_2$ are included. A grid along hyper-radius $\rho$ until 20~bohrs with a grid step of $\Delta \rho$ = 0.08 bohrs was used in the calculation. The largest value of the quantum number $k$ is 4. 
At energies below 50 meV, convergence with respect to the number of channels is poor because of the difficulty to represent the very small reaction probability. To extend the converged results obtained at energies above 50 meV into the low energy region, a WKB approach can be used. Although the tunneling through the potential barrier occurs in the three-dimensional space of internuclear distances of H$_2$D$^-$, for the purpose of using a simplified WKB approach to extrapolate the numercial results below 50~meV, we introduce a generalized tunneling coordinate $x$, which could approximately be viewed as a minimum-energy-path coordinate. Therefore, the WKB formula 
\begin{equation}
   P_{J=0}(E_i) \sim \mathrm{exp} \left( -\frac{2}{\hbar} \int_{w_1}^{w_2} \sqrt{2 \mu (V(x)-E_i)} \, dx \right)\,
  \end{equation}
for the tunneling probability is used, where $w_1$ and $w_2$ are two turning points along $x$. 

In Appendix, we showed that at small $E_i$, the $J$-dependence of the probability behaves as
\begin{align}
 P_J(E_i) \approx P_{J=0}(E_i) \mathrm{exp}\left[-\lambda J(J+1)\right].
 \label{eq:WKB_smallJ}
\end{align}
The sum of reaction probabilities can then be expressed as 
\begin{align}
 \displaystyle\sum_J (2J+1) P_J &\approx P_{J=0} \displaystyle\sum_J (2J+1) \mathrm{exp}\left[-\lambda J(J+1)\right] \nonumber \\
 &\approx \mathrm{exp}\left[AE_i^2 +BE_i +C\right],
 \label{eq:WKB_extra}
\end{align}
where $A,B$, and $C$ are some constant obtained by fitting the result from the ABC program at collision energies between 50 - 80~meV to the WKB formula. 

\section{Results and discussion}


\begin{figure}[ht]
\centering
\includegraphics[width=8cm]{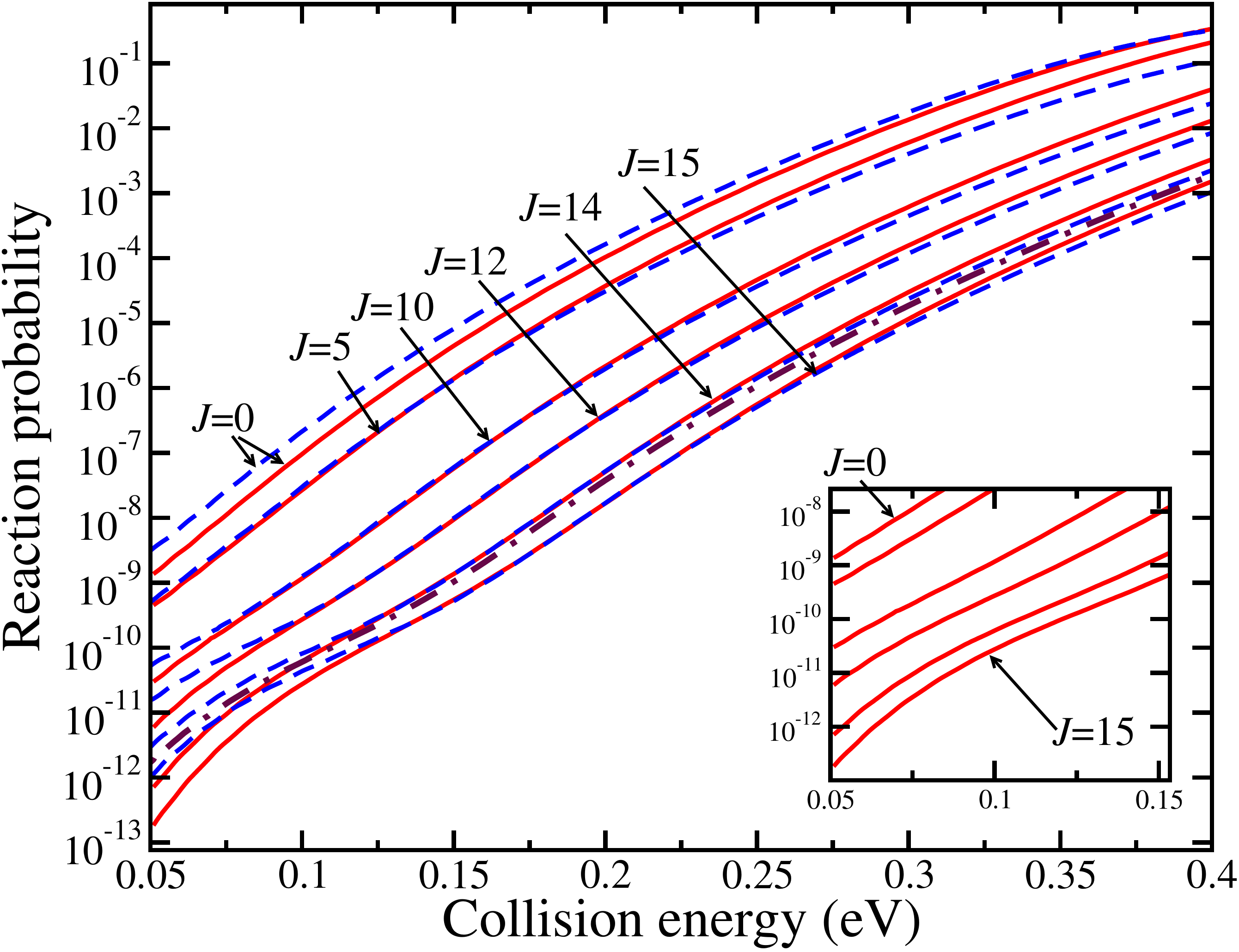}
\caption{(Color online)The blue dashed lines and the red solid lines show the reaction probability of ortho-H$_2$ (with $P=(-1)^J$) and para-H$_2$ in their ground rovibrational states $j=1$ and 0 for the total angular momentum $J$ = 0, 5, 10, 12, 14 and 15. The probabilities decrease with $J$. The dashed-dotted line shows the reaction probability of ortho-H$_2$ with $J$ = 1 and $P$ = 1. The inset shows only the para-H$_2$ curves from the main graph to illustrate the effect of the increasing $J$ on the reaction probability. }
\label{fig:PJ}
\end{figure}

Figure~\ref{fig:PJ} shows the reaction probabilities obtained by solving the Schr\"odinger equation numerically for ortho-H$_2$ and para-H$_2$ in their ground rovibrational states $j=1$ and 0, correspondingly, for different total angular momenta $J$ as a function of collision energy. 
For ortho-H$_2$ with the total parity $P=(-1)^{J+1}$, only the $k=1$ channel can contribute to the reaction. However, the reaction probability for $k=1$ is about three orders of magnitude smaller than for the channel with $P=(-1)^{J}$ and $k=0$ at $J=1$. 
Because the ground-state energy for ortho-H$_2$ is about 15 meV higher than for para-H$_2$, the effective reaction barrier for ortho-H$_2$ is lower and, therefore, its reaction probability is larger at low collision energies, as it is evident in Fig.~\ref{fig:PJ}. At higher collision energies, the difference between the ground-state energies is insignificant and the probabilities are almost equal. But due to the three-fold degeneracy in the entrance channel for ortho-H$_2$, the overall reaction probability for para-H$_2$ is somewhat larger than for ortho-H$_2$ at higher collision energies. 
\begin{figure}[ht]
\centering
\includegraphics[width=8cm]{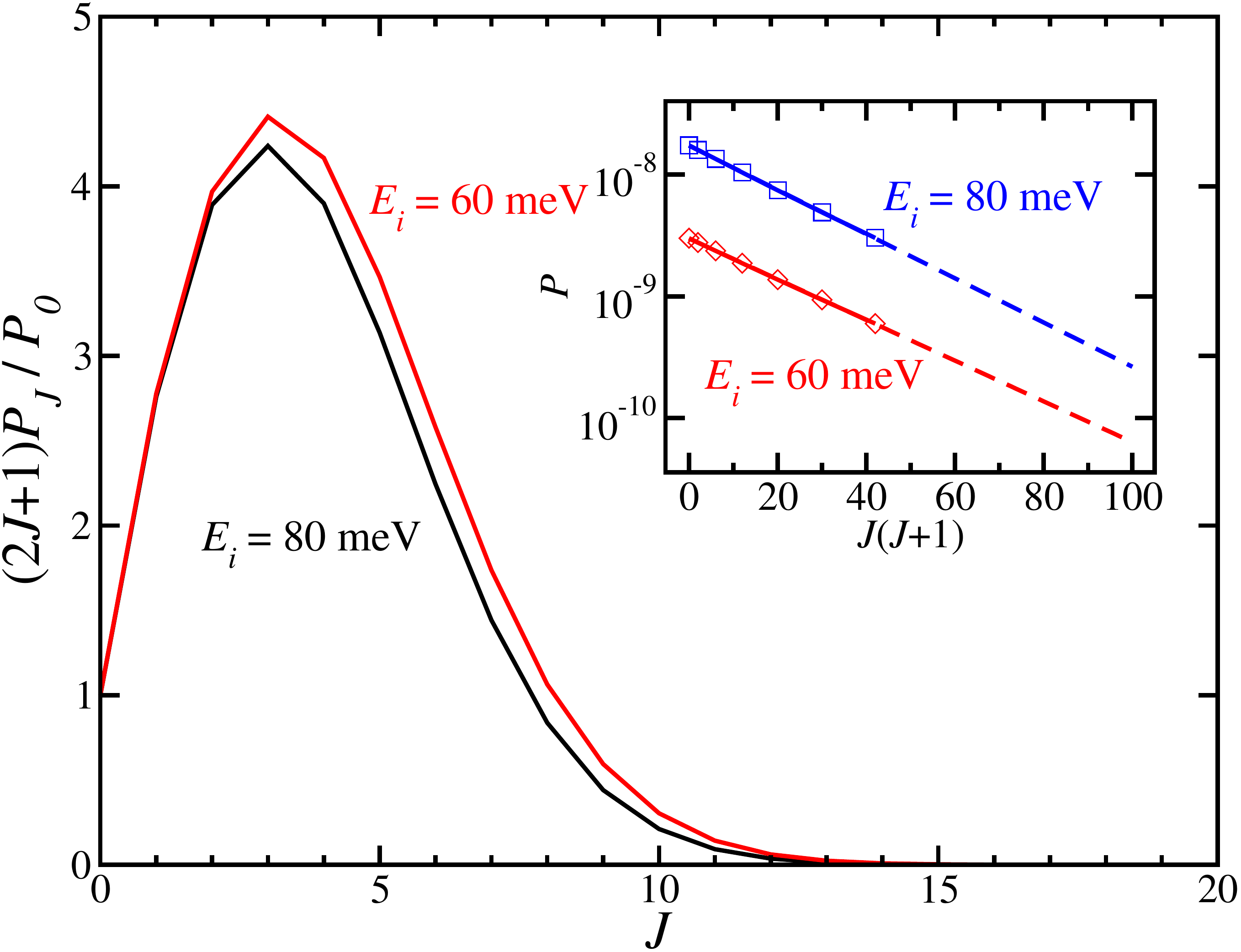}
\caption{(Color online) The figure shows the ratio of $(2J+1)P_J /P_{J=0}$ as as function of $J$ for collision energies $E_i$ = 60~meV (red solid) and 80~meV (black solid) for para-H$_2$. The inset shows the reaction probability at fixed energies of 60~meV (red diamonds) and 80~meV (blue squares) as a function of $J(J+1)$. The dashed lines of the same color in the inset are the linear fit according to Eq.~\eqref{eq:WKB_smallJ}.}
\label{fig:Jtrend}
\end{figure}

At low collision energies, only small values $J$ of the total angular momentum contribute to the total reaction probability as Eq.~\eqref{eq:WKB_smallJ} suggests. The numerical calculations confirm it: The inset of Fig.~\ref{fig:Jtrend} shows that at collision energies of 60~meV and 80~meV, the reaction probability for para-H$_2$ indeed follows the trend of Eq.~\eqref{eq:WKB_smallJ}. The main graph of Fig.~\ref{fig:Jtrend} shows the ratio $(2J+1)P_J /P_{J=0}$ for different $J$. The degeneracy factor $2J+1$ was included in the ratio, because it increases the relative contribution of a particular $J$ into the total cross section. The figure demonstrates that $J$ = 3 contributes the most to the reaction, while for $J >10$ the contributions to the sum of reaction probabilities is small.

\begin{figure}[ht]
\centering
\includegraphics[width=8cm]{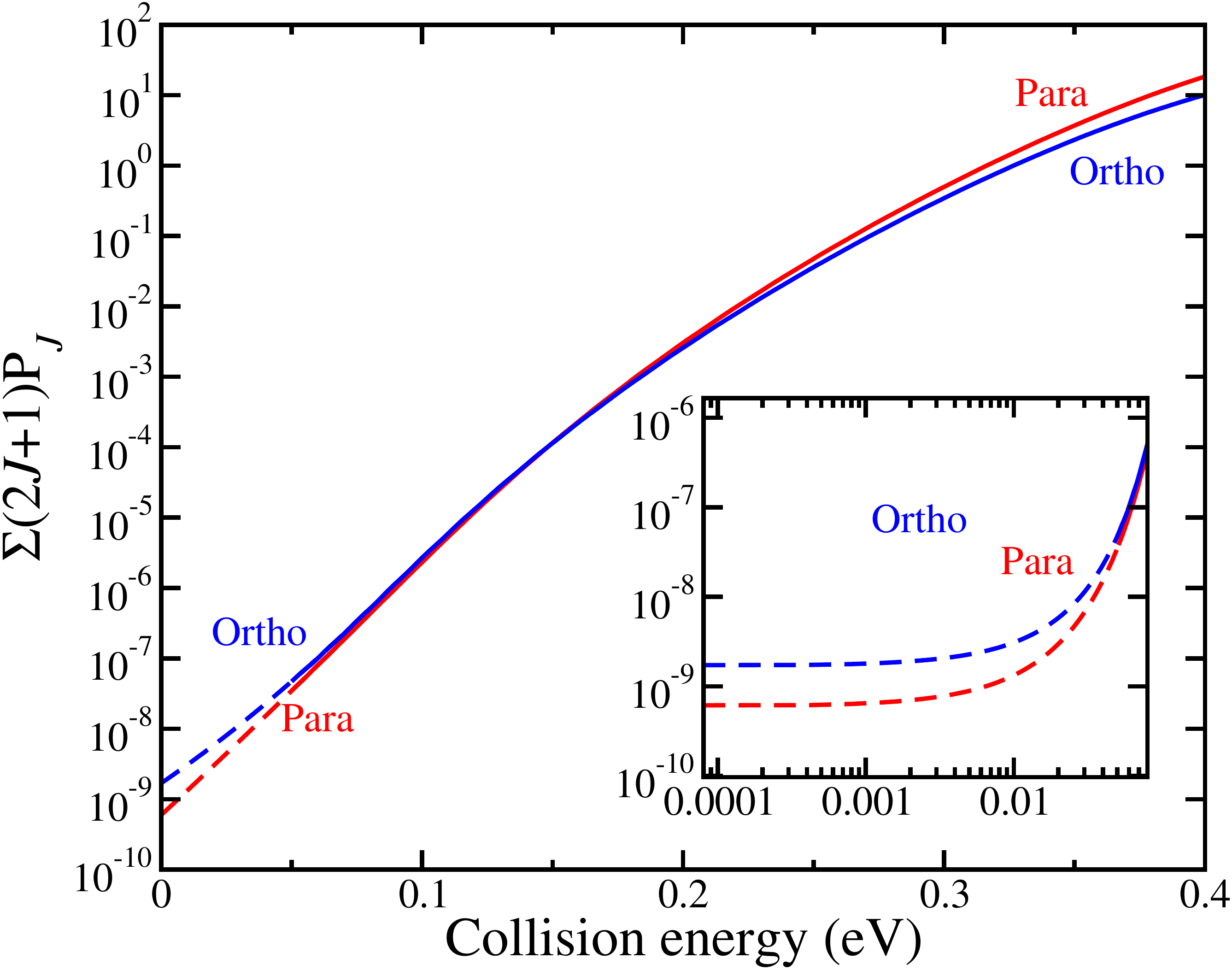}
\caption{(Color online) The figure shows the sum of reaction probabilities $\sum (2J+1) P_J$ for ortho-H$_2$ (blue solid line) and para-H$_2$ (red solid line) as a function of the collision energy. The dashed lines of the same color are the fit to the WKB formula of Eq.~\eqref{eq:WKB_extra}. The uncertainty is within 2\%. The inset is a zoom of the curves in the low energy region.}
\label{fig:Ptotal}
\end{figure}

Figure~\ref{fig:Ptotal} shows the the sum of reaction probabilities $\sum (2J+1) P_J$ for ortho- and para-H$_2$ as a function of the collision energy. The values of $J=0-15$ were included into the sum. In order to extrapolate to the low energy region, we fit the curves obtained numerically to Eq.~\eqref{eq:WKB_extra} using the data points for energies between 50~meV and 80~meV. The uncertainty associated with the fit is within 2\%, which justifies the approximation made in derivation of Eq.~\eqref{eq:WKB_extra}. Furthermore, because the PES has a barrier of 450 meV and due to the H$_2$ zero-point energy  of about 270 meV in the asymptotic H$_2$+D$^-$ region, the effective reaction barrier is lowered to about 180~meV. As the figure indicates, the sum of reaction probabilities becomes significant around that energy and above.

\begin{figure}[ht]
\centering
\includegraphics[width=8cm]{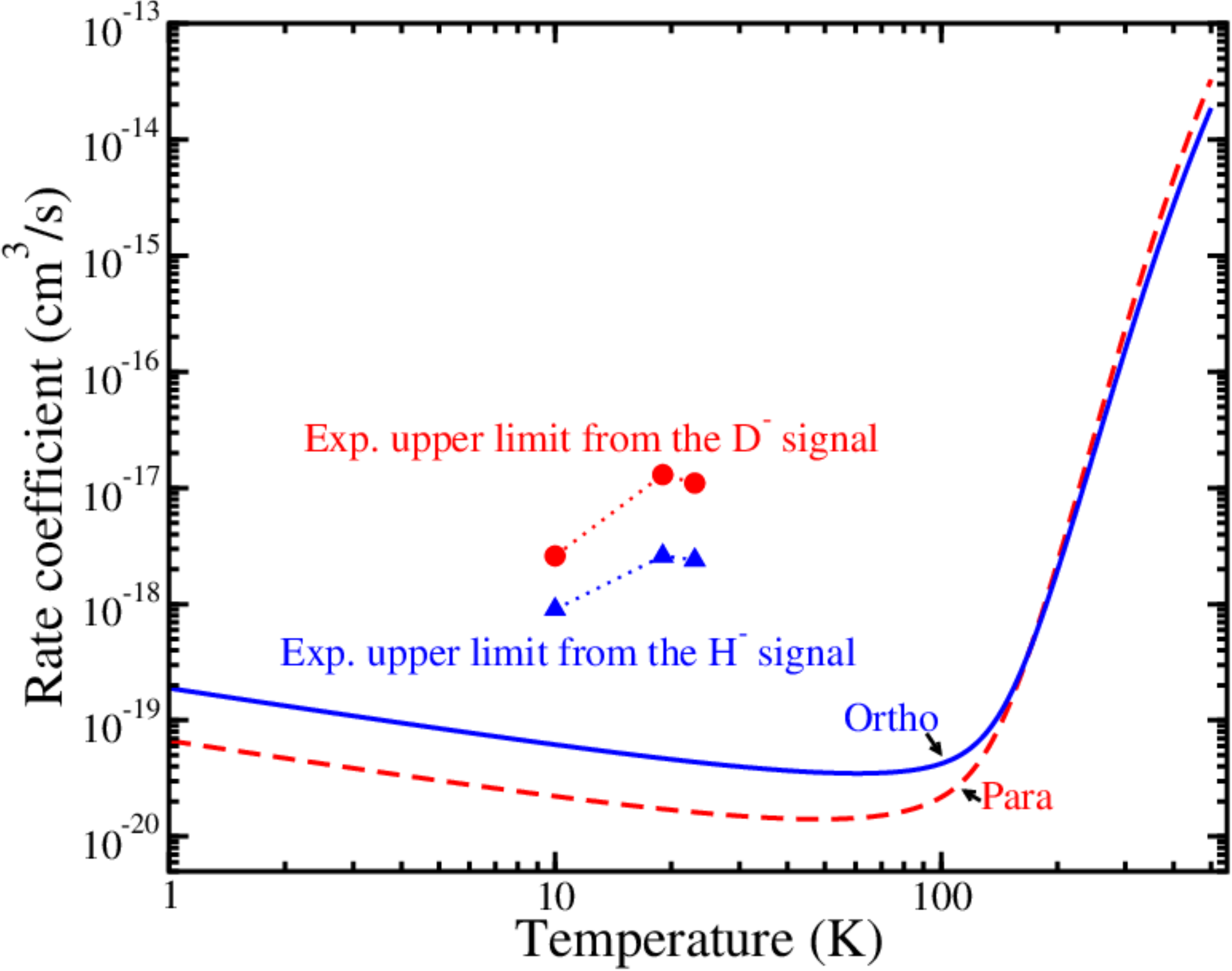}
\caption{(Color online) Thermally averaged rate coefficient for ortho-H$_2$ (blue solid line) and para-H$_2$ (red solid line) as a function of temperature. The red circles and blue triangles represent the experimental upper limits obtained from the D$^-$ signal and from the H$^-$ signal at 10, 19 and 23 K \cite{Endres2017}.}
\label{fig:rate}
\end{figure}

Extrapolated to the low energy region, the thermally averaged rate coefficient is computed using Eq.~\eqref{eq:rate} with the results shown in Fig.~\ref{fig:rate}. The rate coefficients for ortho- and para-H$_2$ are about 3.1 $\times$ 10$^{-20}$ and 1.2 $\times$ 10$^{-20}$ cm$^3$/s, respectively, at temperatures $10-30$~K. The difference between ortho- and para-H$_2$ values at low temperatures is explained by the difference in the ground-state energy and, as a result, by a smaller effective potential barrier for ortho-H$_2$. The thermally averaged rate coefficient for normal hydrogen is closer to ortho-H$_2$ because ortho-H$_2$ has a three times larger statistical weight compared to the para-H$_2$. Therefore, our result is consistent with the experimental upper limit obtained by Endres {\it et al.}

\section{Concluding remarks}
In the study, the thermally-averaged rate coefficient for the nuclei exchange reaction H$_2$+D$^-\to$HD+H$^-$ was computed for temperatures from up to 400~K. In the calculation, the accurate PES \cite{ayouz10} of the system and the ABC code for reactive quantum scattering \cite{Skouteris2000} were used. At low collision energies, a WKB approach was used to extrapolate the results of the fully-quantum approach.  

The obtained thermally-averaged rate coefficient for ortho-H$_2$ is about three times larger than for para-H$_2$ at temperatures below 80~K, while at higher temperatures the coefficients become almost equal. The present theoretical results are about ten times smaller than the experimental upper limit \cite{Endres2017}, suggesting that a further improvement of the sensitivity of the signal in the experiment may lead to the observation of the H$^-$ ions produced from this reaction. Also, the present results suggest that an experiment performed at different temperatures should reveal a strong temperature dependence of the tunneling probabilities at temperatures $70-300$~K.

Future experiments studying collisions of H$_2$ with the H$^-$ isotopes may help to detect eventually the H$^-$ ion in the interstellar space. The ion cannot be detected directly by the absorption or emission spectroscopy, but in collisions with H$_2$ it may form the loosely-bound H$_3^-$ molecule. If detected,  H$_3^-$ could serve as a precursor for H$^-$. The most likely process to form H$_3^-$ is the three-body collisions involving  H$_2$, H$^-$, and a third atom or molecule. In the interstellar space, the third body could be another H$_2$ molecule; in the laboratory, it could be a buffer gas species, such as helium. Therefore, further experiments studying collisions of H$_2$ and H$^-$ are highly desirable.

\section{Acknowledgments}
This work is supported by the National Science Foundation Grant No. PHY-15-06391, the Chateaubriand Fellowship of the Office for Science and Technology of the Embassy of France in the United States. VK acknowledges also a support from the Austrian-American Educational Commission.

\appendix*
\section{Extrapolation using a WKB approach}

\begin{figure}[ht]
\centering
\includegraphics[width=7cm]{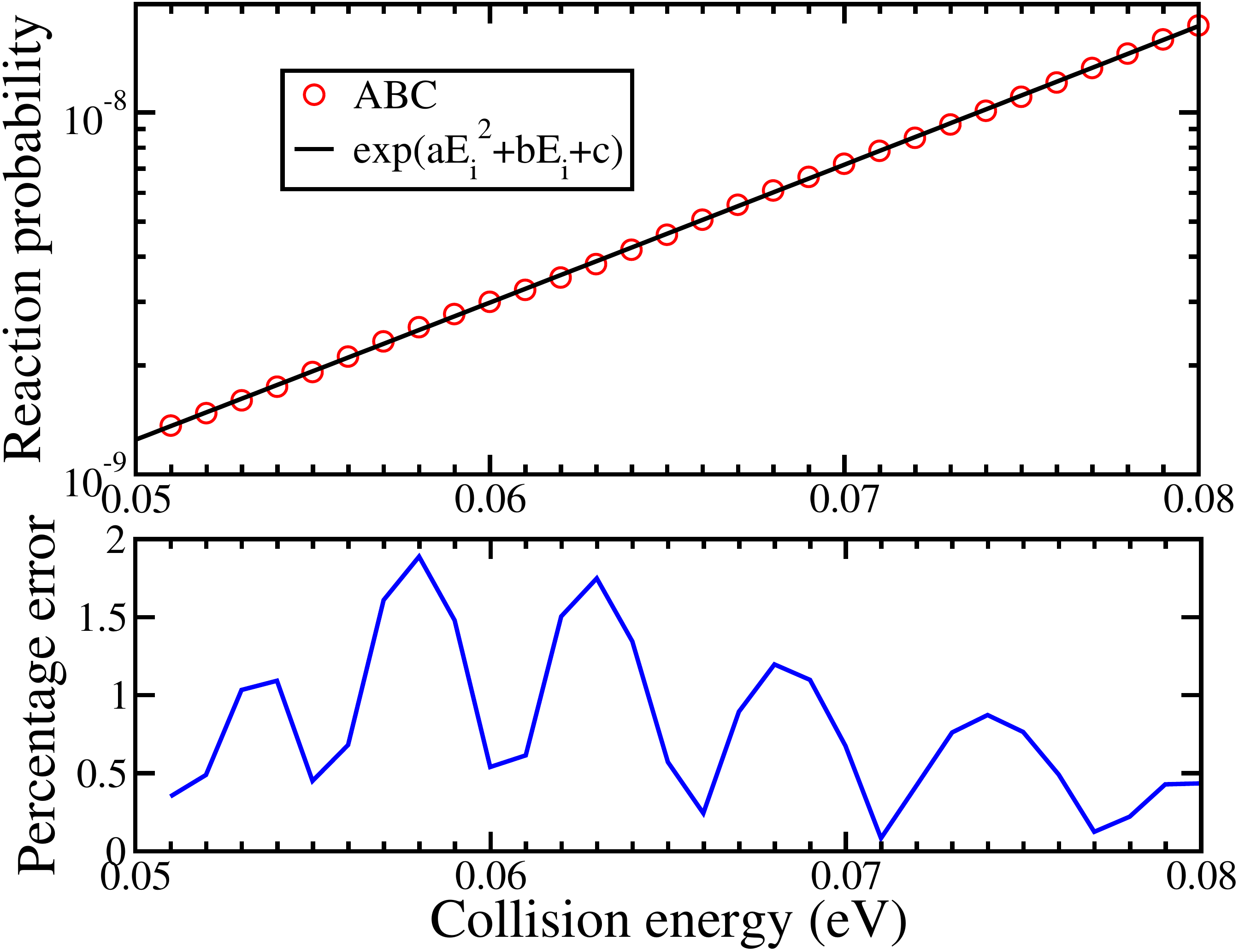}
\caption{(Color online) The upper panel shows the reaction probability of H$_2$(0,0) with the total angular momentum $J=0$ obtained from the fully-quantum approach and the WKB formula of Eq.~\ref{eq:A_WKB_lowE}. The lower panel shows the difference between the two curves in the upper panel.}
\label{fig:WKBfit}
\end{figure}

In the appendix, we derive Eq.~\eqref{eq:WKB_smallJ} and \eqref{eq:WKB_extra} and justify the approximations made above. For collision energies much smaller than the potential barrier, the probability can be expanded in powers of small $E_i$,
\begin{equation}
P_{J=0}(E_i) \approx \mathrm{exp}\left[aE_i^2 +bE_i +c\right]\,
 \label{eq:A_WKB_lowE}
\end{equation}
with
\begin{align*}
 a = \frac{\sqrt{2 \mu}}{4\hbar}  \int_{w_1}^{w_2} \frac{1}{V(x)^{3/2}} \, dx, \, b &= \frac{\sqrt{2 \mu}}{\hbar} \int_{w_1}^{w_2} \frac{1}{\sqrt{V(x)}} \, dx, \\
 c = -\frac{2\sqrt{2 \mu}}{\hbar}  \int_{w_1}^{w_2}  \sqrt{V(x)} \, dx,
\end{align*}
where $V(x)$ is the potential barrier. We assume that at $x=0$, the potential barrier reaches its peak. The quantities $a,b$ and $c$ depend on energy implicitly. To include their energy dependence in Eq.~\eqref{eq:A_WKB_lowE}, one can again expand them in powers of small $E_i$. Because the potential barrier is a very steep function of $x$, increasing the collision energy $E_i$ decreases slightly the width of the barrier, i.e. brings the turning points $w_1$ and $w_2$ closer to each other. It means that $a, b$, and $-c$ are positive quantities, decreasing slowly with $E_i$ if the energies is much smaller than the effective potential barrier of about 180~meV. The matching between the WKB and fully-quantum results is performed near 50-80~meV. To evaluate the uncertainty of the fit, the simple case of H$_2$(0,0) with the total angular momentum $J=0$ was considered.  Figure~\ref{fig:WKBfit} shows the fit using Eq.~\ref{eq:A_WKB_lowE}. The lower panel of the figure demonstrates that the difference between the fit and the fully-quantum result is about 2\% or less, which justifies the validity of the WKB extrapolation.

To account for the dependence of the probability on the total angular momentum $J$, we note that the centrifugal energy $B(x)J(J+1)$ at the top of the barrier at $x=0$ with $B(0)\sim 0.1$~meV at low  $J$ ($\leq 6$) is much smaller than the potential barrier itself. Therefore, we can simply change $E_i$ to $E_i- B(x)J(J+1)$ in Eq~\eqref{eq:A_WKB_lowE}. At $E_i = 80$ meV, $E_i$ is about ten times or more larger than the centrifugal energy. Therefore, the larger contribution from the $J$-dependence of the probability can be accounted for as
\begin{align}
 P_J(E_i) \approx P_{J=0}(E_i) \mathrm{exp}\left[-\lambda J(J+1)\right], \nonumber\, \\
 \lambda = \frac{\sqrt{2 \mu}}{\hbar} \int_{w_1}^{w_2} \frac{B(x)}{\sqrt{V(x)}} \, dx\,.
 \label{eq:A_WKB_smallJ}
\end{align}
For large $J$, the above approximation is not accurate because the centrifugal barrier is significant, but the reaction probabilities are small for $J>6$. (see Figs.~\ref{fig:PJ} and \ref{fig:Jtrend}.) 
Finally, the sum of reaction probabilities can be expressed as
\begin{align}
 \displaystyle\sum_J (2J+1) P_J &\approx P_{J=0} \displaystyle\sum_J (2J+1) \mathrm{exp}\left[-\lambda J(J+1)\right] \nonumber \\
 &\approx \mathrm{exp}\left[AE_i^2 +BE_i +C\right],
 \label{eq:A_WKB_extra}
\end{align}
where we combined exponents in $\mathrm{exp}\left[-\lambda J(J+1)\right]$ and  $P_{J=0}$, and introduced new constants $A,B$, and $C$.


%


\end{document}